\documentclass[manuscript]{emulateapj}
\usepackage{amsmath}

\shorttitle{FRB from Clusters}
\shortauthors{Fialkov, Loeb \& Lorimer}
\usepackage{color}

\begin{document}

\title{Enhanced Rates of Fast Radio Bursts from Galaxy Clusters}

\author{Anastasia Fialkov$^{1}$, Abraham Loeb$^{1}$, Duncan R.~Lorimer$^{2,3}$}
\affil{\small $^{1}$Harvard-Smithsonian Center for Astrophysics, Institute for Theory and Computation, 60 Garden Street, Cambridge, MA 02138, USA}
\affil{\small $^{2}$Center for Gravitational Waves and Cosmology, West Virginia University, Chestnut Ridge Research Building, Morgantown,
WV 26505, USA
}
\affil{\small $^{3}$Department of Physics and Astronomy, West Virginia University, Morgantown, WV 26506, USA
}
\email{anastasia.fialkov@cfa.harvard.edu}

\begin{abstract}
Fast Radio Bursts (FRBs) have so far been detected serendipitously across the sky. We consider the possible enhancement in the FRB rate in the direction of galaxy clusters, and compare the predicted rate from a large sample of galaxy clusters to the expected cosmological mean rate. We show that clusters offer better prospects for a blind survey if the faint end of the FRB luminosity function is steep. We find that for a telescope with a beam of $\sim 1$~deg$^2$, the best targets would be either nearby clusters such as Virgo or clusters at intermediate cosmological distances of few hundred Mpc, which offer maximal number of galaxies per beam. 
We identify several galaxy clusters which have a significant excess FRB yield compared to the cosmic mean. The two most promising candidates are the Virgo cluster containing 1598 galaxies and located 16.5~Mpc away and S34 cluster which  contains 3175 galaxies and is located at a distance of 486~Mpc. 
\end{abstract}


\section{Introduction}

Fast radio bursts (FRBs) are rapid transients detected in the $\sim 0.7-1.8$~GHz frequency range and  characterized by a few millisecond duration. Since  the discovery of the first FRB in 2007 \citep{Lorimer:2007}, 23 additional  bursts were observed by several radio telescopes in different regions of the sky \citep[][see the online FRB catalog\footnote{http://www.frbcat.org} for more details on the detected events]{Keane:2011, Thornton:2013, Burke-Spolaor:2014, Petroff:2015, Ravi:2015, Champion:2016, Keane:2016,  Ravi:2016,  Petroff:2017, Masui:2015, Caleb:2017, Bannister:2017}. The repetitive nature of one of the bursts, FRB121102, allowed its localization to a few arcminutes and the identification of the host galaxy at a redshift 0.2 \citep{Chatterjee:2017, Tendulkar:2017}. This discovery demonstrated that at least some FRBs are of cosmological origin. 

FRBs located at large cosmological distances can be used as probes of both their host environment and the intergalactic medium along the line of sight. As an FRB propagates through the ionized intergalactic medium, its pulse is dispersed in a frequency-dependent manner.  The dispersion measure (DM) is proportional to the integrated electron column along the line of sight in (units of pc~cm$^{-−3}$) which can be directly related to the redshift of the source after the contributions of the host galaxy and of the Milky Way are subtracted out. If FRBs exist prior to the Epoch of Reionization, their DM can constrain the reionization history and measure the total optical depth with sub-percent accuracy \citep{Fialkov:2016}. Surveying  the population of FRBs could, therefore, not only reveal their origin, but also improve our understanding of cosmic history.

Up to now, FRBs have been discovered serendipitously across the sky. However, observational effort is on the way to perform more focused FRB searches and pin down the nature of these sources. Future surveys include the Canadian Hydrogen Intensity Mapping Experiment (CHIME), which  is expected to have 125 mJy flux limit  in the $400-800$ MHz frequency range and a large collecting area \citep{Newburgh:2014, Rajwade:2017},  as well as targeted searches with the Green Bank Telescope (GBT) at 1.4 GHz. Future facilities such as the Square Kilometer Array (SKA) are predicted to  detect many more of these events  \citep{Fialkov:2016, Fialkov:2017}. 

The origin of FRBs is still a mystery and it is unclear what are the properties, progenitors and  host galaxies of these transients \citep[e.g.,][]{Houde:2017,Metzger:2017,Beloborodov:2017,Cordes:2016}. Recently, Macquart \& Ekers (2017) examined the population of FRBs and found that current data is weakly inconsistent with flat luminosity function and implies only a very weak constraint on the slope of the integrated number counts to be $ < -−1.3$ with the most likely value being $ -−2.6^{+0.4}_{-0.6}$. 

In this paper we study the possible enhancement in the FRB rate through  observations of dense environments such as rich galaxy clusters. To bracket the large uncertainty we consider different scenarios varying the nature of the progenitors and the luminosity function of FRB. The paper is organized as follows. In Section \ref{Sc:Model} we outline our model assuming that the population of FRBs is of  cosmological origin. In  Section \ref{Sc:cataloges} we consider the Virgo cluster as a prototype and estimate the rate and distribution of FRBs from the cluster center using a public catalog of galaxies  \citep{Kim:2014}. In Section \ref{Sc:results} we apply the formalism to a large sample of clusters from  public galaxy cluster catalogs of the Sloan Digital Sky Survey \citep[SDSS][]{Einasto:2007, Liivamagi:2012}. We summarize our conclusions in Section \ref{Sc:Conc}.

\section{Cosmological population of FRBs}

\label{Sc:Model}

The expected  rate and spatial distribution of FRBs strongly depends on their origin. Even under the assumption of a cosmological origin,  there is a large variety of possible progenitors of FRBs. Since the host galaxy population is not yet constrained by observations,  we consider two different scenarios in our modeling, assuming that FRBs are produced by either old or young stars. In addition, we consider two different shapes of the FRB luminosity function and vary the luminosity of the faintest events. Our cosmological models are summarized in the first  Column of Table 1. We assume no repetitions of FRBs in our calculation.

\begin{table*}
\begin{center}
\begin{tabular}{  lc  c c cccc }
Model  & $R_{\rm int}/\dot N_{4}$ &   $N^{\rm Cosm}_{{\rm all}~z,~1~{\rm deg}^2}/\dot N_{4}$    &  $N^{\rm Virgo}_{{\rm max},~1~{\rm deg}^2}/\dot N_{4}$  & $N^{\rm Cosm}_{V_{\rm Virgo}}/\dot N_{4}$  &  $N^{\rm Virgo}_{V_{\rm Virgo}}/ \dot N_{4}$ \\
\hline
\#1 M*, SC &  314 M$_{\odot}^{-1}$ &  1.6 & 0.099 & $8.4\times 10^{-4}$ &   1.58 \\
\#2 M*, Sch$^{\alpha = -2}_{inst}$ &  $8.9\times 10^{11}$ M$_{\odot}^{-1}$& 0.92 &  $6.5\times 10^{3}$ & 55.3 &$1.04\times 10^5$   \\
\#3 M*, Sch$^{\alpha = -2}_{obs}$    &    $6.0\times 10^{3}$ M$_{\odot}^{-1}$& 1.06 &  1.89  & 0.016 & 30.2 \\
\#4 SFR, SC & 0.0012 yr$^{-1}$ &  0.45& 0.0525 &  $1.9\times 10^{-4}$ & 0.899 \\
\#5 SFR, Sch$^{\alpha = -2}_{inst}$ & $8.2\times 10^6$  yr$^{-1}$& 0.58 &  $8.6\times 10^3$  &  1.1 & $1.5\times 10^5$  \\
\#6 SFR, Sch$^{\alpha = -2}_{obs}$  & 0.048  yr$^{-1}$ &0.59&  2.18 &  0.008 & 37.4  \\
\end{tabular}
\caption{Summary of model predictions assuming a flat spectrum for FRBs and $S_{\rm lim}=$~30 Jy. {\bf Column 1:} model description, low-luminosity cutoff $L_{\rm min}$ is determined either by the instrument ($S_{\rm min}^{\rm inst}$) or by the observed FRB with the lowest intrinsic luminosity ($S_{\rm min}^{\rm obs}$); {\bf Column 2:} the inferred FRB normalization per galaxy in units of  $\dot N_{4} = \dot N_{\rm obs}/10^4$ where $\dot N_{\rm obs}\sim 10^3-10^5$ sky$^{-1}$~day$^{-1}$ is the observed rate; {\bf Column 3:}  average number of FRBs per 1 deg$^2$~yr$^{-1}$  integrated over the entire redshift range out to $z = 10$ in units of $\dot N_{4}$;  {\bf Column 4:}  peak FRB rate from Virgo per year in a 1 deg$^2$~beam  in units of   $\dot N_{4}$.  {\bf Column 5:} average number of FRBs per year in units of  $\dot N_{4}$ from a random patch of the sky of the virial volume of Virgo ($4\pi R_{\rm vir}^3 /3$, $R_{\rm vir} =1.72$) located at the redshift of Virgo $z = 0.002$ (16.5 Mpc). {\bf Column 6:} same number as in Column 5 but for the real distribution of galaxies in Virgo extracted from the online Virgo catalog \citep{Kim:2014}.   }
\end{center}
\label{Tab1}
\end{table*}

In the  first scenario FRBs are produced in star forming regions and trace the population of newly born (massive) stars. In this case,  the rate of FRBs in each individual galaxy would be proportional to its star formation rate (SFR), $\dot N_{1}  = R^{\rm int}_{\rm SFR}\times$SFR, where SFR is in units of M$_\odot$ yr$^{-1}$. $ R^{\rm int}_{\rm SFR}$ is the normalization constant (in units of M$_\odot ^{-1}$, yielding the FRB rate in units of yr$^{-1}$). When considering a cosmological population of galaxies, we adopt the SFR derived by \citet{Behroozi:2013}. This model, based on observations across a wide range of stellar masses $M_* \sim 10^7-10^{12}$ M$_\odot$ and redshifts\footnote{In our analysis we extrapolate the model out to redshift 10. However, high redshift FRBs do not have any impact on the results presented in this paper.} ($z=0-8$), provides the star formation rate as a function of dark matter halo masses $\left(M_h\right)$ and redshift. 

The second scenario  is that FRBs are produced by old progenitors. In this case, FRB rate (in units of yr$^{-1}$) scales as the total stellar mass, $M_*$, and is  $\dot N_1 =  R^{\rm int}_{*}M_*/M_{\rm Virgo}$ with $R^{\rm int}_*$ being the normalization constant in units of yr$^{-1}$. In the context of clusters, we normalize the total stellar mass by the mass of the Virgo galaxy cluster, $M_{\rm Virgo} = 1.2\times 10^{15}$ M$_\odot$. Stellar mass can be related to the host halo mass \citep[e.g.,][]{Mashian:2016} via the star formation efficiency which we also adopt from the work by  \citet{Behroozi:2013}.

The FRB rate from  a large cosmological volume $V$ is obtained by integrating over the entire population of star forming halos in it. The number of halos in each mass bin $\Delta M_h$ is $\Delta M_h dn/dM_h$  per comoving Mpc$^3$, and can be derived from the Press--Schechter formalism \citep{Press:1974}, or more accurately from the Sheth--Tormen mass function \citep{Sheth:1999} which was calibrated against numerical simulations. The rate of FRBs in units of sky$^{-1}$ yr$^{-1}$ observed at redshift $z=0$ from the entire cosmological galaxy population  is thus
\begin{equation}
 \dot N_{\rm FRB} = \int_VdV \int_{M_h} dM_h \frac{dn}{dM_h}\frac{\dot N_1}{(1+z)},
 \label{Eq:Nfrb}
\end{equation}
where $V$  is the comoving volume and we integrate over host halo mass.  The redshift factor $(1+z)^{-1}$ accounts for cosmological time dilation. 

To bracket the large uncertainty in the FRB luminosity functions we consider two different scenarios:\\
(i) FRBs are standard candles (SC) of the same peak luminosity $\nu L_\nu = 2.8\times 10^{43}$ erg s$^{-1}$, which corresponds to the mean intrinsic luminosity of the observed FRBs. To derive this value we used the online FRB catalog. For each event, the intrinsic isotropic luminosity can be derived based of the reported peak flux $S_{peak}$ and the redshift estimated from the DM, with $L_\nu = 4\pi D_L^2 S_{\rm peak}(1+z)^{-1}$ where $D_L$ is the luminosity distance. We then multiply by $\nu =1$ GHz, the typical frequency at which FRBs are observed, to get $\nu L_\nu$ and calculate the mean value across the  ensemble of the observed FRBs.\\
(ii)  FRBs have a Schechter (Sch) luminosity function 
\begin{displaymath}
\frac{dn}{dL_\nu} = \left(\frac{L_\nu}{L_{\nu *}}\right)^{-\alpha}\exp\left[-\frac{L_\nu}{L_{\nu *}}\right]
\end{displaymath}
with $\nu L_{\nu *} = 2.8 \times 10^{43}$ erg s$^{-1}$ and the faint-end slope of $\alpha =  -2$. This is the steepest slope for which the luminosity density of a cosmological population converges, and this slope is broadly consistent with current observational constraints \citep{Mccart:2017}.

An additional free parameter in the case of a Schechter luminosity function is the low luminosity cutoff, $L_{\nu, min}$, the lowest luminosity of FRBs. In popular theoretical models  FRBs, are launched by young magnetars \citep{Cordes:2016, Beloborodov:2017, Metzger:2017};  however, FRBs appear to be $\mathcal{O}\left(10^{10}\right)$ times brighter than the typical magnetars found in our vicinity \citep{Maoz:2017}.  To allow for the wide range of possibilities  we, therefore, consider two cases: (1) $L_{\rm min}$ is set to be the luminosity of the intrinsically faintest observed FRB, $L_{\nu, {\rm min}} = L_{\nu, {\rm min}}^{\rm obs}$, namely  FRB010621 with $\nu L_{\nu}^{\rm peak} = 5.1\times 10^{41}$ erg s$^{-1}$; (2) FRBs can be as faint as the galactic magnetars resulting in $L_{\nu, {\rm min}} = \left(L_{\nu, *}/10^{10}\right)$. In the latter case, it is the telescope sensitivity,  $S_{\rm lim}$, which sets the lower limit on the flux of observed events.

The number of events detected by a given radio observatory depends on  several factors. To be detectable, the flux of a redshifted burst  should be above the sensitivity limit of the telescope, $S_{\rm lim}$, and, if the burst has a limited frequency band, it should fall within the sensitivity band of the telescope. Here, for simplicity, we assume a flat spectrum (i.e., flux being independent of observing frequency) and set the flux limit to $S_{\rm lim}=30$ Jy having in mind a wide field survey with a small radio telescope. One such experiment is currently in operation at the
Green Bank Observatory  and makes use of the
20~m antenna there to carry out
searches for FRBs at 1.4~GHz (Golpayegani et al. in prep.). This system
has $S_{\rm lim}=30$~Jy over
a 1~deg$^2$ field of view.

To calibrate each cosmological model we compare the expected  rate of FRBs from Eq. (\ref{Eq:Nfrb}) to the observational constraint which yields  $\dot N_{\rm obs} \sim 10^3-10^5$~FRBs~sky$^{-1}$ day$^{-1}$ at $z<1$ and $S_{\rm lim}\ge 1$ Jy  \citep[e.g.,][]{Keane:2015, Nicholl:2017, Law:2017}. The results for normalization in each case are shown in the second Column of Table 1, in units of $\dot N_4$ sky$^{-1}$ yr$^{-1}$, where $\dot N_4 \equiv 10^{-4} \dot N_{\rm obs}$ and $\dot N_{\rm obs}$ is in units of $\rm{sky}^{-1}~ \rm{day}^{-1}$. For the Schechter luminosity function with the low-luminosity cutoff being set  by telescope sensitivity, there is no way to constrain the faint end of the population and many faint events can occur per galaxy. This  explains the very high relative normalization and expected number counts from nearby clustered environments.  
Using our cosmological model and  integrating over the entire redshift range out to $z = 10$, we then compute the mean FRB rate expected from a solid angle of  1 deg$^2$  per year (Column 3 of Table 1) and observed by a telescope with $S_{\rm lim} = 30$ Jy.  

\section{FRB Rate from the Virgo Cluster}
\label{Sc:cataloges}

Next, we explore the expected FRB rates from clustered environments  and compare the predicted numbers to the cosmological mean derived above. As a proof of concept we focus on the nearby Virgo cluster. Using the online Virgo catalog \citep{Kim:2014} which lists cluster members and the luminosity of each galaxy in every SDSS band, we infer stellar masses and SFRs for each galaxy in the cluster and estimate the expected number of FRBs for the actual distribution of galaxies.

\subsection{Stellar mass}
\label{SMS}
Stellar masses can be derived for individual Virgo galaxies by using standard mass-luminosity relations.  To derive total stellar mass we follow \citet{Bernardi:2010}. The mass-luminosity relation at redshift $z = 0$  is given as a function of $(g-r)_0$ colors 
\begin{displaymath}
\log_{10}\left(M_*/L_r\right) = 1.097\left(g-r\right)_0+z_p,
\end{displaymath}
where $z_p$ depends on the initial mass function (IMF) and, following \citet{Bernardi:2010},  we set $z_p=-0.406$ \citep[Chabrier IMF,][]{Bernardi:2010}. The magnitude in the r-band (which provides the luminosity in the r-band, $L_r$) is calculated\footnote{For the r-band the correction to the AB system is negligible and $r_{\rm AB}\approx r_{\rm SDSS}$.} as $M_r = r_{AB}-5\left[\log(D_{\rm Virgo,~pc})-1\right]$. Stellar mass is then calculated from
\begin{equation}
\log_{10}M_* = 1.097(g-r)_0-0.406-0.4 (M_r-4.67)
\label{Eq:Mstar}
\end{equation}
and $(g-r)_0$ is extracted from the catalog.

\subsection{Star Formation Rate}
The SFR in star forming galaxies follows a well known characteristic relation with the stellar mass \citep[e.g.,][]{Brinchmann:2004}  referred to as the  main sequence of galaxies \citep[e.g.,][]{Noeske:2007} and  parametrized as 
\begin{equation}
\log_{10}(\rm {SFR}) = a\log_{10}\left(M_* \right)+b.
\label{SFRM}
\end{equation}
To compute SFR for the Virgo galaxies, we apply the aperture-free SFR-M$_*$ relation \citep{Duarte:2017} and use $M_*$ obtained in the Section \ref{SMS} above. \citet{Duarte:2017} derived the total SFR for $\sim$210,000  SDSS star-forming galaxies using an empirically based aperture correction of the measured H$\alpha$ fluxes which have been extinction-corrected. The SFR$-M_*$  relation has been obtained  in six redshift bins, over the redshift range $0.005 < z < 0.22$ with $a = 0.935$ and $b =-9.208$. We use these values of $a$ and $b$ in  Eq. (\ref{SFRM}) to estimate the SFR of each galaxy in the  Virgo cluster. 

\subsection{Expected FRB Rate}

We use the derived $M_*$ and SFR, along with the number of galaxies extracted from the catalog, to calculate the expected rate of FRBs from the entire Virgo cluster. As we see from Table 1, the largest effect on the observed FRB rate from nearby sources (e.g., galaxies in Virgo) is that of the luminosity function, while the nature of the hosts (young versus old stars) has a stronger effect on the cosmological background rate. With the real spacial distribution of galaxies in Virgo \citep{Kim:2014}, we infer the expected rate of FRBs per each beam of 1 deg$^2$ and show the resulting sky distribution   in Figure 1 with the assumptions of model \#6  from Table 1. The few bright regions on this map indicate the optimal spots to target in a  future search for FRBs in Virgo. 

\begin{figure}
\begin{center}\includegraphics[width=3.4in]{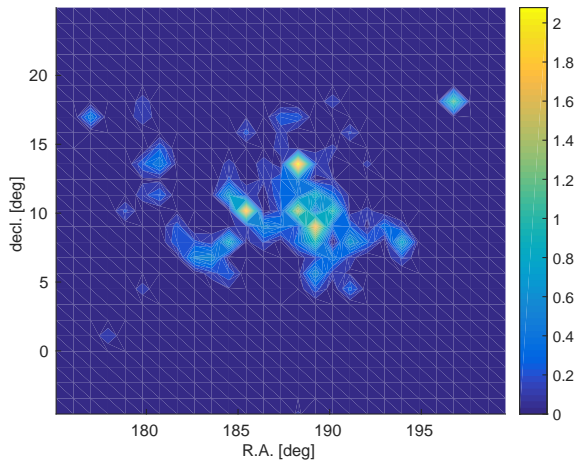}
\caption{FRB rates from Virgo in deg$^{-2}$ yr$^{-1}$ assuming $\dot N_{\rm obs} = 10^4$ sky$^{-1}$ day$^{-1}$   for  the model \#6  from Table \ref{Tab1}. Coordinates of the region with the highest FRB rate within Virgo are R.A. = 188.28 and Dec. = 13.58. } 
\end{center}
\label{Fig:FRBVirgo}
\end{figure}

For a wide-band spectrum of FRBs (similar to the flat spectrum assumed here) observing the clustered environment is  beneficial only if the faint-end slope of the luminosity function is steep \citep[such as suggested by current observations][]{Mccart:2017}. This can be seen by comparing Column 3 to Column 4 in Table \ref{Tab1} for the rates within a 1 deg$^2$. If the population of faint FRBs is significant, the rate from clusters will exceed the cosmological mean by factor of a few in models \#3 and 6 and by few orders of magnitude in models \#2 and 5. On the other hand, if FRBs are standard candles \citep[models \#1 and 4, mildly inconsistent with observations][]{Mccart:2017}, dense nearby clusters such as Virgo would only contribute $\sim 10\%$ of the total observed FRB rate. Thus, clusters offer a new way to test the faint end of the luminosity function of FRBs.

Spectrum of FRBs also plays a role. If FRBs are narrow-band \citep[e.g., similar to FRB121102][]{Law:2017}, only FRBs from a bounded redshift range fall within the telescope band. In this case, the FRB rate from clustered environments might exceed the mean cosmological rate even if they are standard candles. We demonstrate this by comparing the FRB rates for the virial volume of Virgo inhabited  by a mean cosmological population of galaxies (Column 5) to the rate generated by a real distribution of galaxies in the cluster. For the  scenarios under consideration, the total FRB yield is more than $1000$ times larger from the cluster than from a random field of the same virial size. 

\section{FRB from Galaxy Clusters}
\label{Sc:results} 

Next, we apply the formalism outlined above to a larger sample of galaxy clusters located at comoving distances out to $\sim 800$ Mpc,  using two different catalogs, namely the 2dF catalog \citep{Einasto:2007}  and the SDSS-DR7 sample \citep{Liivamagi:2012}.   The catalogs provide information on the number of galaxies within virial radius of each cluster. Assuming that the number of FRBs scales as the number of galaxies, we estimate the FRB rate per each individual cluster by simply re-scaling the number counts from Virgo. The expected intrinsic rate from a cluster is thus  $\dot N_{\rm FRB}^{\rm cl} = \dot N_{\rm FRB}^{\rm Virgo}\times N_{\rm gal}^{\rm cl}/ N_{\rm gal}^{\rm Virgo}$. FRB rate per cluster and the average FRB rate per beam for each cluster are shown in Figure 2 for each one of the considered models. To calculate the FRB rate per beam we divide the total FRB rate from the virial volume of each cluster by max$\left[A_{\rm eff},1~ \rm{deg}^2\right]$ with $A_{\rm eff}$ being the effective area of the cluster. The rate from clusters is compared to the cosmological mean background (horizontal lines). In Figure 2 we also show the rate for Virgo (diamonds) and Coma \citep[squares, extracted from the SDSS catalog of][]{Liivamagi:2012} clusters for comparison.

\begin{figure*}
\begin{center}
\includegraphics[width=3.4in]{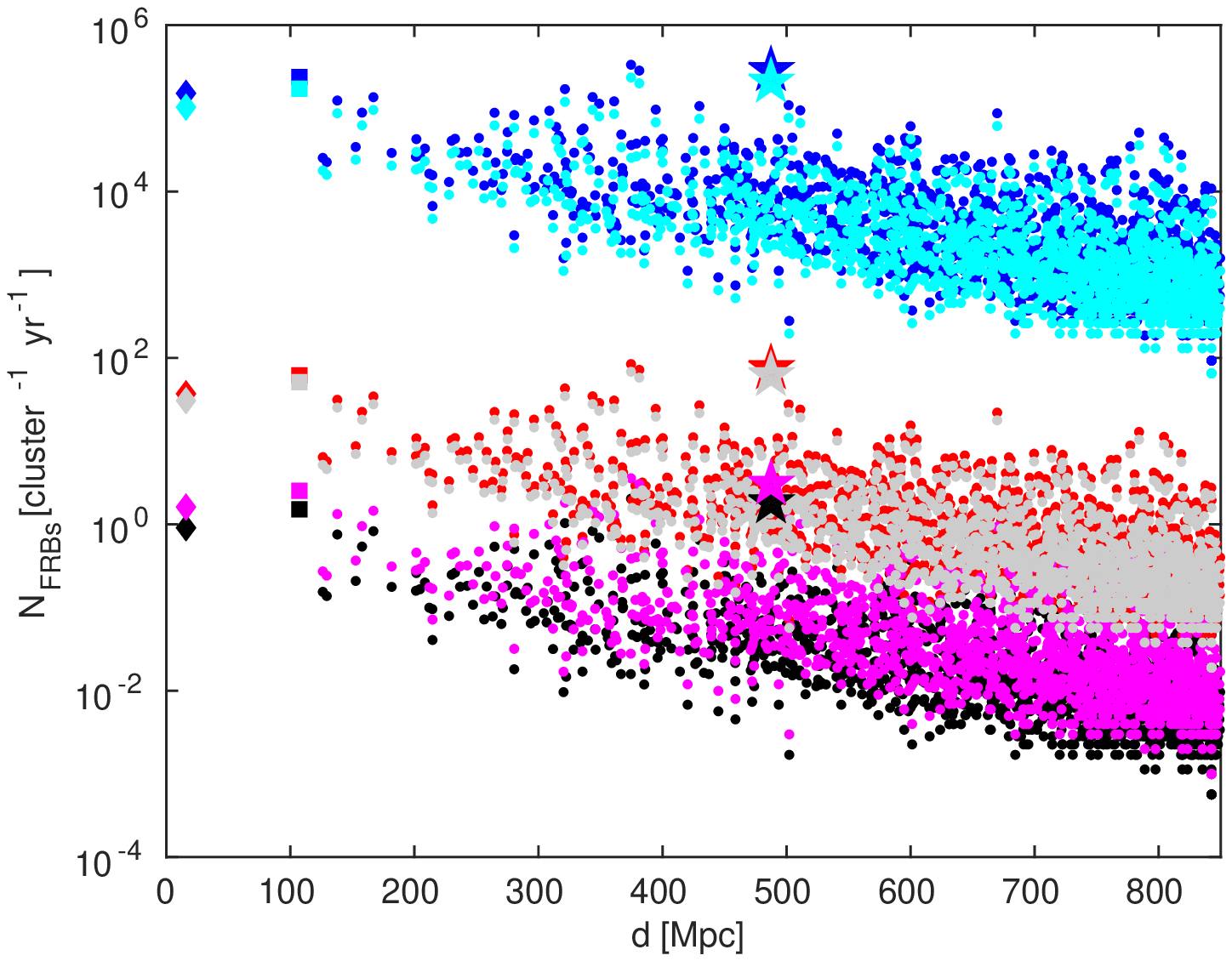}\includegraphics[width=3.4in]{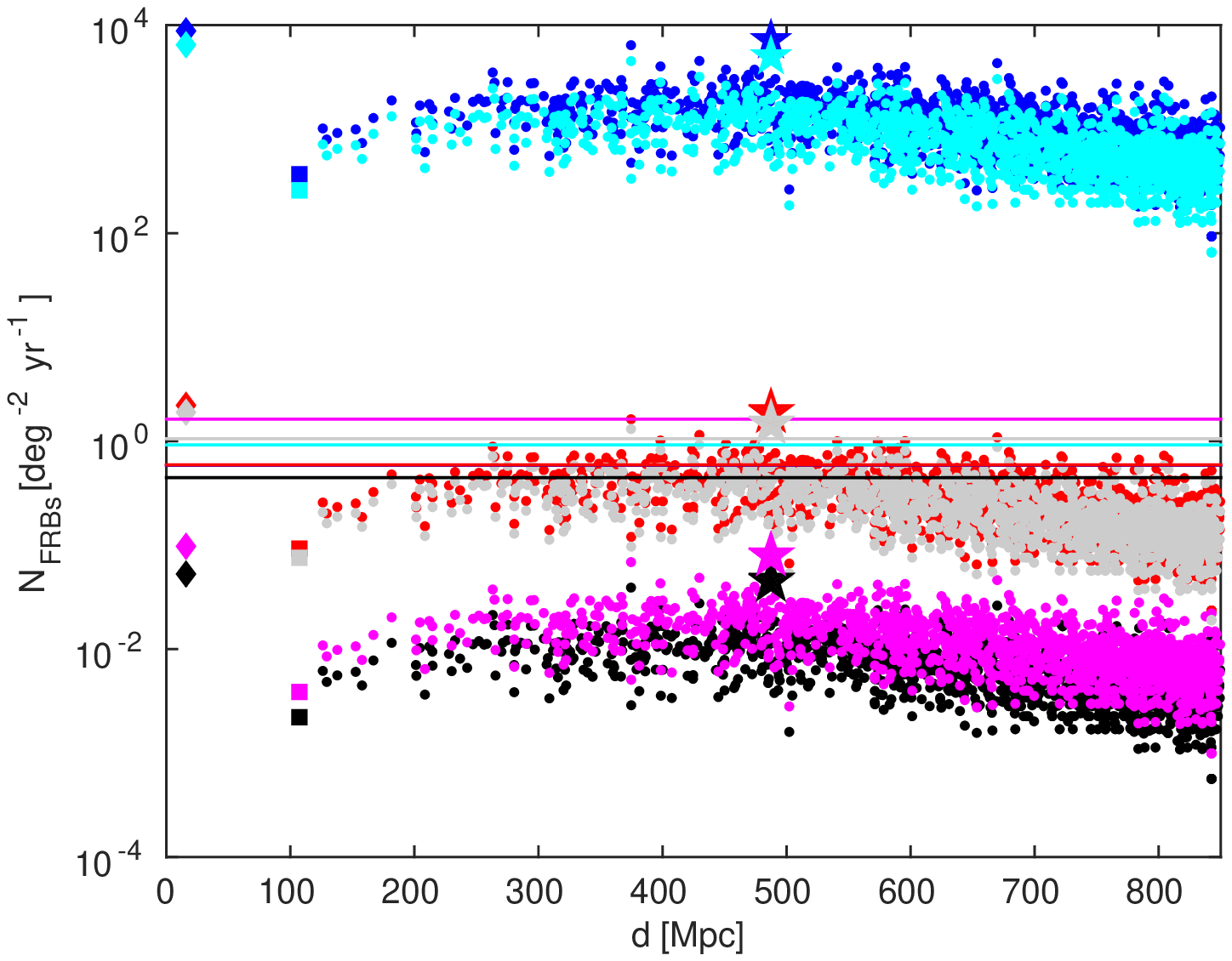}
\caption{FRB rate in each  cluster (left) and the maximal rate  per 1 deg$^2$  beam (right) for all the considered models: \#1 (magenta), \#2 (cyan), \#3 (green), \#4 (black),  \#5 (blue),  \#6 (red). Solid horizontal lines on the right-hand side panels correspond to the cosmological mean estimate. Here we assume a normalization of 10$^4$ [sky$^{-1}$ day$^{-1}$] FRBs. Diamonds indicate total (left) and maximal (right) number counts from Virgo; square markers denote same numbers estimated for the Coma cluster (from the SDSS data); stars mark the cluster with the highest $\dot N_{FRB}^{cl}$. } 
\end{center}
\label{Fig:Clusters}
\end{figure*}

As in the case of Virgo, the largest uncertainty in the predicted FRB rate is introduce by the poor understanding of the luminosity function; while the  nature of the progenitors has only a minor effect. If FRBs are standard candles (models \#1 and 4), their contribution is negligible compared to the cosmological background; while if the faint population is significant (models \#2 and 5),  $\dot N_{FRB}^{cl}$ exceeds the cosmological contribution by few orders of magnitude. 
  
An interesting case is of our models \#3 and 6 where the minimal luminosity is matched to the faintest observed FRB. In this case only part of the clusters have high FRB yield, and the best candidates for the targeted FRB searches with an instrument of 1 deg$^2$ beam are galaxy clusters located at intermediate cosmological distances, $\sim 300-700$ Mpc (Figure 2). This is because the number of galaxies per the beam is optimal at such distances.

\begin{figure*}
\begin{center}
\includegraphics[width=3.4in]{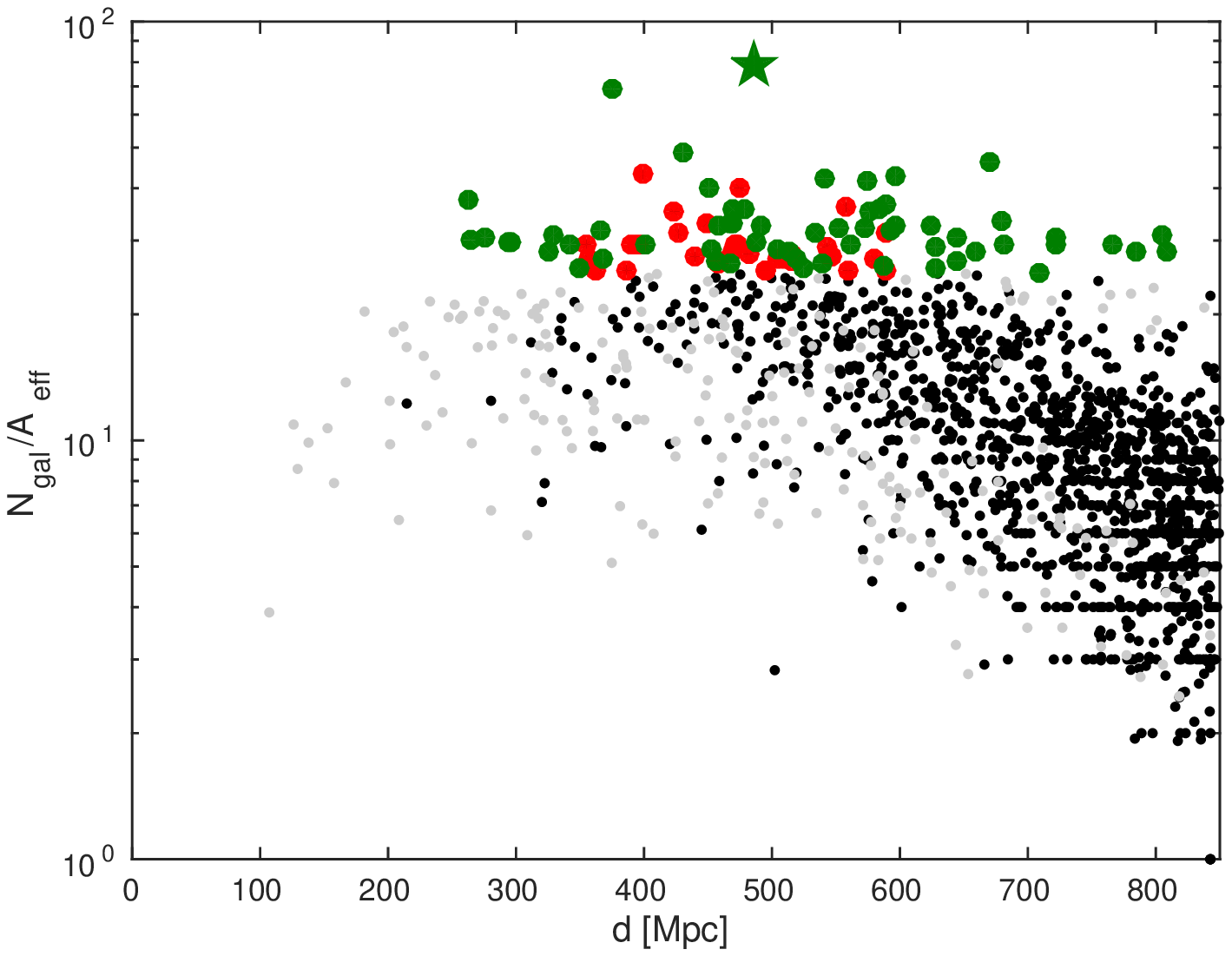}\includegraphics[width=3.4in]{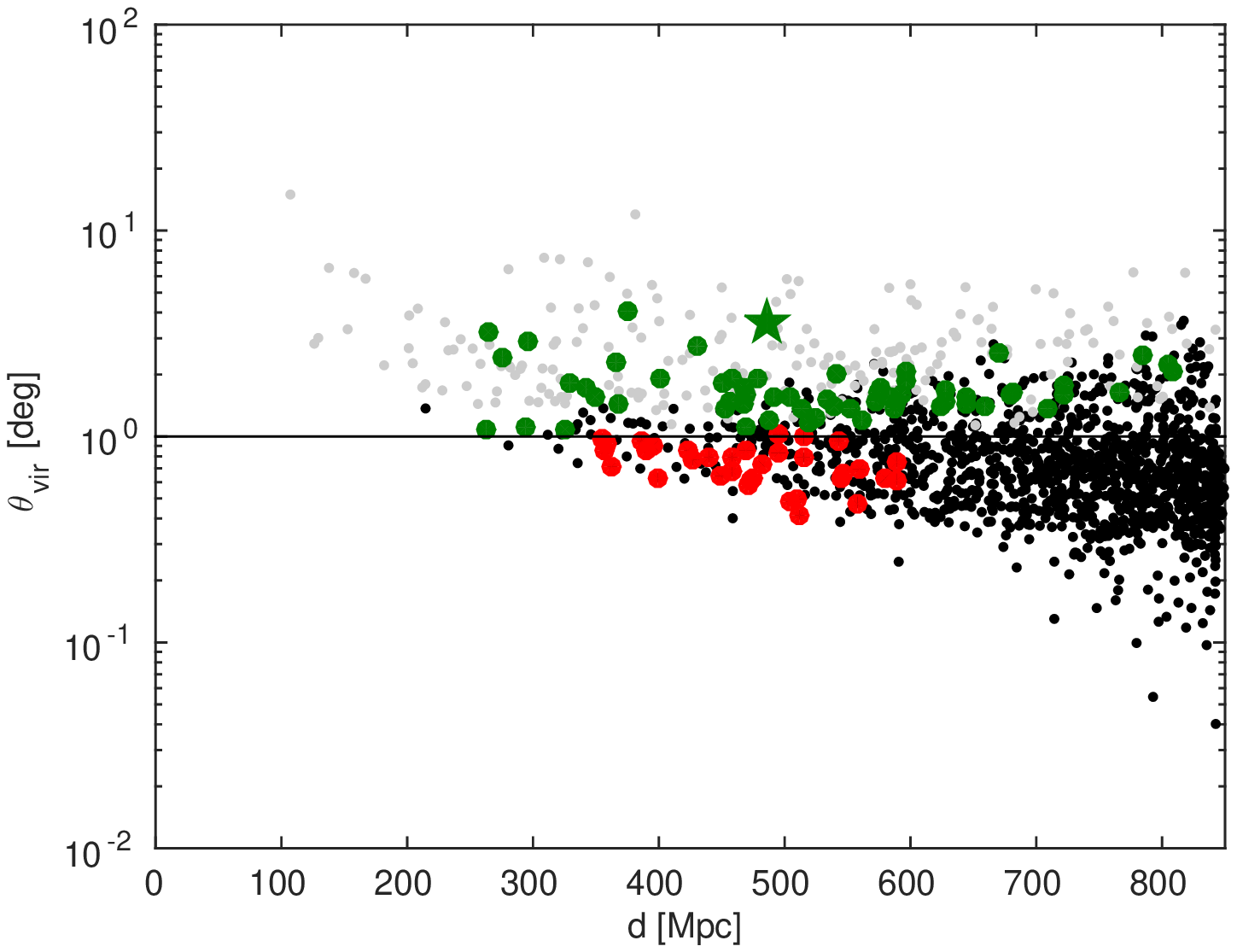}
\caption{Number of galaxies per effective area of the cluster (left) and  the angular radius of each cluster (right) shown for all clusters from the 2dF survey \citep{Einasto:2007}  and SDSS DR7 \citep{Liivamagi:2012}. Circles mark clusters with $\dot N_{FRB}^{cl}$ higher the cosmic mean for model \#6. Stars denote the cluster with the highest FRB yield. Dots show clusters with $\dot N_{FRB}^{cl}<\dot N_{FRB}^{cosm}$.  We plot rich   ($N_{gal}>100$, green and grey) and poor  ($N_{gal}<100$, red and black) clusters.  The black horizontal line (right) refers to the beam size (1 deg). } 
\end{center}
\label{Fig:Clusters2}
\end{figure*}

Adopting our model \#6 as a reference, we examine for which of the SDSS clusters $\dot N_{\rm FRB}^{\rm cl}$ exceeds the cosmological background. The number of galaxies per effective area of the cluster and the angular size of each cluster compared to the beam size are shown in Figure 3 where we mark (circles and stars) clusters with $\dot N_{\rm FRB}^{\rm cl}$ above the cosmic mean.  

It is evident that the clusters yielding elevated FRB rate  are those with the largest number of galaxies per effective area. We find that there are two types of clusters that contribute: (i) rich clusters which host large number of galaxies ($N_{gal}>100$, green circles in Figure 3), and (ii) poor clusters ($N_{gal}<100$, red circles in Figure 3) of angular size comparable to the telescope resolution. We find the best candidate for the targeted FRB search to produce 3.1 more FRBs than the background with the assumptions of model \#6 (and 1.4 for \#3). This candidate (marked with a star in Figures 2 and 3) is a rich cluster containing 3175 galaxies, located  at a distance of 486 Mpc towards R.A. = 9.8$^\circ$ and Dec.= -28.9$^\circ$. We give details of this cluster, as well as an additional 15 candidates (including Virgo) in Table 2. The close proximity of Virgo relative to the other clusters
we have considered so far still elevates it to the highest ranking in Table 2,
despite the fact that it is not fully sampled by a 1~deg$^2$ beam. A targeted
survey with a wider field instrument, such as the Australian Square Kilometre Array
Pathfinder \citep[ASKAP;][]{Bannister:2017} which has a 30~deg$^2$ field of view
would provide a dramatic increase in these rates.

\begin{table*}
\begin{center}
\begin{tabular}{  ll   r r rrcc }
Rank & Cluster name   & $N_{\rm gal}$  & $D$\footnote{Note that in the catalogs \citep{Einasto:2007, Liivamagi:2012} the distances are given in [Mpc/h] units. We use h= 0.6704 \citep{Planck:2016a} for conversion.}   & R.A.  &  Dec.  & Boost & $N^{\rm cl}_{1~{\rm deg}^2}/\dot N_{4}$ \\
     &   & & Mpc & \multicolumn{2}{c}{deg} &  & \\
\hline
 1 & Virgo \citep{Kim:2014} & 1598 & 16.5 & 188.28\footnote{We quote R.A. and Dec. of the region with the highest FRB rate within Virgo.}& 13.58 & 3.69 & 2.18 \\ 
 2 & S 34 \citep{Einasto:2007} & 3175 & 486 & 9.86 & --28.94  & 3.12 & 1.85 \\
 3 & N 512 \citep{Einasto:2007} & 3591 & 375 & 194.71 & --1.74  & 2.74 & 1.62 \\
 4 & N 13 \citep{Einasto:2007} & 1145 & 430 & 152.01 & 0.57 & 1.93 & 1.15 \\
 5 & S 217 \citep{Einasto:2007} & 938 & 670  & 334.75 & --34.76  & 1.84 & 1.09 \\
 6 & 235+017+0089 \citep{Liivamagi:2012}& 54 & 398  & 235.16 & 18.14 & 1.71 & 1.01 \\
 7 &N 99 \citep{Einasto:2007}  & 472 & 596 & 177.62 & --0.60 & 1.69 & 1.00\\
 8 & S 10 \citep{Einasto:2007} & 535 & 541 & 3.02 & --27.42 & 1.68 & 1.00\\
 9 &  N 37 \citep{Einasto:2007} & 359 & 574 & 160.34 & --5.90 & 1.66 & 0.99 \\
 10 & N 76 \citep{Einasto:2007} & 420 & 451 & 170.64 & 0.45 & 1.60 & 0.95\\
 11 & 133+000+0108 \citep{Liivamagi:2012}& 50 & 474 & 133.69 & 0.75 & 1.59 & 0.94 \\
12 & 223+018+0059 \citep{Liivamagi:2012} &  138 & 263 & 223.47 &  18.82 & 1.50 &  0.89 \\
13 & N 136 \citep{Einasto:2007} & 251 &  590 & 190.10 & --4.44 & 1.45 & 0.86 \\ 
14 & 147+007+0127 \citep{Liivamagi:2012}& 36 & 558 & 147.28 & 7.19 & 1.43 &  0.85 \\
15 &S 126 \citep{Einasto:2007} & 291 & 469 & 34.36 & --29.43 & 1.42 & 0.84 \\
16 &N 170 \citep{Einasto:2007} & 415 & 478  & 200.94 & 1.08 & 1.42 & 0.84 \\
\end{tabular}
\caption{Top 16 search candidates.  {\bf Column 1:} number; {\bf Column 2:} catalog and cluster name; {\bf Column 3:} number of galaxy members; {\bf Column 4:} distance [Mpc];
{\bf Column 5:} R.A.; {\bf Column 6:} Dec.; {\bf Column 7:} ratio between the predicted FRB rate per 1 deg $^2$ beam from the cluster to the cosmic mean with the assumptions of model \#6; {\bf Column 8:}   the predicted FRB rate per year in a 1 deg $^2$ beam in units of   $\dot N_{4}$ with the assumptions of model \#6.  }
\end{center}
\label{Tab2}
\end{table*}

\section{Conclusions}
\label{Sc:Conc}

We have considered the contribution of galaxy clusters to the total FRB rate. For targeted FRB searches with radio telescope beam sizes of 1 deg$^2$ and sensitivity limit $S_{lim} = 30$ Jy, observing either nearby clusters (such as Virgo) or clusters at intermediate cosmological distances (a few hundred Mpc) is the best strategy. We find that the predicted rate from clusters strongly depends on the FRB luminosity function and in particular on its  faint end slope, whereas  the nature of hosts (young versus old stars) has a less significant impact. If the FRB luminosity function has a steep faint-end slope, clusters will provide a dominant contribution to the observed events, while if the faint-end slope is shallow the main contribution will be from the  cosmological background. Comparing the rates within a beam which includes a cluster versus the field will thus constrain the number of faint FRBs and the luminosity of the population. 
This analysis makes definitive predictions in the form of a number of promising
galaxy cluster targets (see Table 2) for future  observational campaigns with radio telescopes. Although our analysis here has focused on instruments with 1~deg$^2$ 
beams as its basic unit, wider field instruments with comparable sensitivity 
for example ASKAP will be able to play a significant role in constraining the
FRB luminosity function through deep stairs at nearby rich clusters.

\acknowledgments
We thank J.~Guillochon and K.~Bannister for useful discussions. 
This work was supported in part by the Breakthrough Prize Foundation and Harvard's Black Hole Initiative. DRL is supported by
NSF AST-1516958.

\end{document}